\documentclass[pra,twocolumn,showpacs,letterpaper,showpacs,superscriptaddress]{revtex4}
\usepackage{amsfonts}
\usepackage{amsmath}
\usepackage{amssymb}
\usepackage{graphicx}
\usepackage{latexsym,color,dcolumn,bm,epsfig,subfigure,pstricks-add}

\newcommand{\la}{\langle}
\newcommand{\ra}{\rangle}
\newcommand{\be}{\begin{equation}}
\newcommand{\ee}{\end{equation}}
\newcommand{\bea}{\begin{eqnarray}}
\newcommand{\eea}{\end{eqnarray}}
\newcommand{\om}{\omega}
\newcommand{\Om}{\Omega}

\newcommand{\pa}{\partial}
\newcommand{\br}{{\bf r}}
\newcommand{\bx}{\hat{\bf x}}
\newcommand{\bbx}{{\bf x}}
\newcommand{\bp}{\hat{\bf p}}
\newcommand{\bb}{\hat{\bf b}}
\newcommand{\bK}{\hat{\bf K}}
\newcommand{\hK}{{\hat{K}}}
\newcommand{\hC}{\hat{C}}
\newcommand{\hg}{\hat{g}}
\newcommand{\eps}{\epsilon}

\newcommand{\bE}{\hat{\bf E}}
\newcommand{\bF}{\hat{\bf F}}
\newcommand{\bH}{\hat{\bf H}}
\newcommand{\bA}{\hat{\bf A}}
\newcommand{\bD}{\hat{\bf D}}
\newcommand{\bB}{\hat{\bf B}}
\newcommand{\bP}{\hat{\bf P}}
\newcommand{\bbE}{{\bf E}}
\newcommand{\bbD}{{\bf D}}
\newcommand{\bbH}{{\bf H}}

\newcommand{\bk}{{\bf k}}

\newcommand{\bek}{{\bf e}_{{\bf k}\lambda}}
\preprint{ }

\begin{document}

\title{Electromagnetic Energy, Absorption, and Casimir Forces. I. Uniform Dielectric Media in Thermal Equilibrium}

\author{F. S. S. Rosa}
\affiliation{Theoretical Division, Los Alamos National Laboratory, Los Alamos, NM 87545, USA}

\author{D. A. R. Dalvit}
\affiliation{Theoretical Division, Los Alamos National Laboratory, Los Alamos, NM 87545, USA}

\author{P. W. Milonni}
\affiliation{Theoretical Division, Los Alamos National Laboratory, Los Alamos, NM 87545, USA} 

\begin{abstract}
The derivation of Casimir forces between dielectrics can be simplified by ignoring absorption, calculating energy changes due to displacements of the dielectrics, and only then admitting absorption by allowing permittivities to be complex. As a first step towards a better understanding of this situation we consider in this paper the model of a dielectric as a collection of oscillators, each of which is coupled to a reservoir giving rise to damping and Langevin
forces on the oscillators and a noise polarization acting as a source of a fluctuating electromagnetic field in the dielectric. The model leads naturally to expressions for the quantized electric and magnetic fields that are consistent with those obtained in approaches that diagonalize the coupled system of  oscillators for the dielectric medium, the reservoir, and the electromagnetic field. It also results in a fluctuation-dissipation relation between the noise polarization and the imaginary part of the permittivity; comparison with the Rytov fluctuation-dissipation
relation employed in the well-known Lifshitz theory for the van der Waals (or Casimir) force shows that the
Lifshitz theory is actually a {\sl classical} stochastic electrodynamical theory. The approximate classical expression for the energy density in a band of frequencies at which absorption in a dielectric is negligible is shown to be {\sl exact} as a spectral thermal equilibrium expectation value in the quantum-electrodynamical theory.
Our main result is the derivation of an expression for the QED energy density of a uniform dispersive, absorbing media in thermal equilibrium. 
The spectral density of the energy is found to have the same form with or without absorption. We also show how the fluctuation-dissipation theorem ensures a detailed balance of energy exchange between the (absorbing) medium, the reservoir and the EM field in thermal equilibrium.
\end{abstract} 

\pacs{42.50.-p, 03.70.+k, }

\maketitle

\section{Introduction}

Based on the assumption of an electromagnetic (EM) energy ${1\over 2}\hbar\om$ per mode
of angular frequency $\om$ at zero temperature, Casimir \cite{casimir} famously showed that there is an attractive force between two uncharged, perfectly conducting plates. Lifshitz \cite{lifshitz} generalized the theory to the case of two thick dielectric slabs in thermal equilibrium by calculating the stress tensor for the fluctuating field in a vacuum between the slabs. Casimir's original method involving changes in zero-point field energy was later extended to dielectrics by van Kampen {\sl et al.} \cite{vankampen} in the ``nonretarded" case of small separations, and by others for arbitrary separations \cite{others}.

Derivations of the Lifshitz formula that invoke changes in zero-point energy begin by assuming real
dielectric permittivities. After an integral over frequencies for the force as a function of the distance separating the dielectrics is obtained, the permittivities are allowed to be complex functions, analytic in the
upper half of the complex frequency plane as required by causality. This allows an analytic continuation to an integral involving only purely imaginary frequencies, at which the permittivities are purely real, and the resulting expression is
equivalent to that of Lifshitz, who requires complex permittivities (absorption) through 
the fluctuation-dissipation relation. In this paper we take a first
step towards a better understanding of why this approach leads in the end to results that are
equivalent to those obtained by Lifshitz-type approaches, which are not based on calculations of energy changes
and which account explicitly for absorption via a fluctuation-dissipation relation. 

We begin in the following section by revisiting the classical theory of electromagnetic energy density for quasi-monochromatic fields in
a dispersive, absorbing medium. At frequencies $\om$ at which absorption is negligible the
classical expression for the average energy density is \cite{dispersion_no_absorption}
\be
\label{NoAbsorption1}
\overline{u}(\br,\om)= \frac{1}{16 \pi} \left[  
\frac{d (\omega \epsilon_R)}{d \omega}  |{\bf E}_{\om}({\bf r})|^2  + 
 \frac{d (\omega \mu_R)}{d \omega}  |{\bf H}_{\om}({\bf r})|^2\right],
\ee
where $\eps_R$ and $\mu_R$ are respectively the (real) permittivity and magnetic permeability
at frequency $\om$, $\bbE_{\om}(\br)$ and $\bbH_{\om}(\br)$ are the electric and magnetic fields at $\om$ and the bar over $u({\bf r},\om)$ indicates a time average (see section II.A). In the following section it is also shown that, within a band of frequencies for which absorption is negligible, $u(\br,\om)$ gives {\it exactly} the
spectral energy density as long as the fields at different frequencies 
undergo uncorrelated fluctuations. In the quantum-electrodynamical (QED) theory in which there are quantum
field fluctuations, this holds for the expectation value of the thermal equilibrium energy, and in
particular for the zero-point energy. 
More interesting for our purposes, however, is the fact that these expectation
values in QED have the same form {\it with or without absorption} in uniform media. 

This paper is organized as follows. In Section \ref{sec:classical} we review some aspects of the classical theory
of electromagnetic energy density in dispersive, absorbing media. Equation (\ref{NoAbsorption1}) is shown 
to give the total energy density of the field and the polarizable particles of a purely dielectric medium ($\mu=1$)
at frequencies for which dissipation is negligible. Whereas it gives approximately the total energy density in the
case of quasi-monochromatic fields, it gives exactly the average total spectral energy density within a band of frequencies
in which different frequencies undergo uncorrelated fluctuations, provided that absorption can be ignored in this band.
We review in Section \ref{sec:zpe} the total {\it quantum}-electrodynamical thermal equilibrium energy density for a dispersive magnetodielectric medium in which absorption is negligible. In Section \ref{sec:micro} we treat in some detail the quantum theory of a dielectric medium modeled as a collection of polarizable material oscillators, allowing for dissipation by coupling each of these oscillators to a reservoir and deriving the fluctuation-dissipation
relation for the noise polarization arising from quantum fluctuations of the bath oscillators. We argue, based on this
derivation, that the Lifshitz theory employing such a fluctuation-dissipation relation is in fact a {\it classical} stochastic electrodynamical theory. Proceeding as in the
Lifshitz theory in which the noise polarization acts as a source of a fluctuating electromagnetic field, we obtain expressions for the quantized electromagnetic field in a dissipative medium; these quantized
fields have the same form as in the Huttner-Barnett theory in which the Hamiltonian for the coupled system of oscillators for the dielectric medium, the reservoir, and the electromagnetic field is diagonalized \cite{huttnerbarnett}. Adding
each contribution to the total energy density, including that from the reservoir, we derive (for the first time to our knowledge), 
the expression (\ref{energyabsorption}) for the QED energy density of a uniform dispersive, absorbing medium in thermal equilibrium. 
We show in particular that it has exactly the same form as obtained in Section \ref{sec:zpe} when dissipation is neglected, and is affected by dissipation only through the dependence on dissipation of the real part of the refractive index. In
Section \ref{sec:summ} we summarize our conclusions and discuss briefly how they apply to two examples: the Einstein $A$ coefficient for spontaneous emission and the van der Waals interaction of atoms embedded in a dissipative dielectric medium. 

This is intended to be the first of two papers dealing with electromagnetic energy in dispersive, dissipative media. 
The approach described here will be extended in a forthcoming paper on effects of dissipation when Casimir effects (presence of material boundaries) are calculated following an approach based on zero-point energy \cite{paperII}.


\section{Classical electromagnetic energy density in dispersive, absorbing media}\label{sec:classical}

The problem of defining and calculating an electromagnetic energy density for the classical EM field in dispersive and absorbing media was investigated in detail by Barash and Ginzburg \cite{Barash1976, GinzburgPlasma}. Here we review some of the main points and difficulties associated with this problem. The classical expression for electromagnetic energy density can be derived from the Poynting theorem,
\begin{equation}
- \nabla \cdot {\bf S} = \frac{1}{4 \pi}  \left( {\bf E} \cdot \frac{\partial {\bf D}}{\partial t}  + {\bf H} 
\cdot \frac{\partial {\bf B}}{\partial t} \right) ,
\label{Poynting}
\end{equation}
where ${\bf S} = c \left( {\bf E} \times {\bf H}\right)  / 4 \pi$ is the Poynting vector 
in the conventional notation. We assume that the constitutive relations connecting ${\bf E}$, ${\bf D}$, ${\bf B}$ and 
${\bf H}$ are linear, isotropic and spatially local, so that, writing
\bea
{\bf E}({\bf r},t)&=& \int_{-\infty}^{\infty} d\omega {\bf E}({\bf r},\omega)e^{-i\omega t},
 \nonumber \\
{\bf H}({\bf r},t)&=& \int_{-\infty}^{\infty}d\omega {\bf H}( {\bf r},\omega) e^{-i\omega t},
\eea
we have
\bea
\label{ConstRel}
{\bf D}({\bf r},t)&=& \int_{-\infty}^{\infty} d\omega \epsilon({\bf r},\omega) {\bf E}({\bf r},\omega)e^{-i\omega t},
 \nonumber \\
{\bf B}({\bf r},t)&=& \int_{-\infty}^{\infty}d\omega \mu({\bf r},\omega) {\bf H}( {\bf r},\omega) e^{-i\omega t}.   
\eea
We write the (complex) electric permittivity $\eps$ and magnetic permeability $\mu$ in terms of real and imaginary 
parts: $\epsilon({\bf r},\omega) = \epsilon_R({\bf r},\omega) + i \epsilon_I({\bf r},\omega)$ and $\mu({\bf r},\omega) = \mu_R({\bf r},\omega) + i \mu_I({\bf r},\omega)$. At this point we should stress that we consider only passive media throughout this work, meaning that  $\eps_I(\omega), \mu_I(\omega) > 0$ for all frequencies. The well-known relations
\begin{eqnarray}
\label{RefRels}
&&{\bf E}({\bf r}, -\omega) = {\bf E}^*({\bf r}, \omega) \;\; , \;\; {\bf B}({\bf r}, -\omega) = {\bf B}^*({\bf r}, \omega) , \nonumber \\
&&\epsilon({\bf r}, -\omega) = \epsilon^*({\bf r}, \omega) \;\; , \;\; \mu({\bf r},-\omega) = \mu^*({\bf r}, \omega) 
\end{eqnarray}
follow from the reality of ${\bf E}$ and ${\bf B}$. Then
\be 
\label{Poynting2}
\int_{-\infty}^tdt'(-\nabla\cdot{\bf S})=\mathcal{W}_E+\mathcal{W}_H,
\ee
with
\begin{widetext}
\be
\mathcal{W}_E(\br,t) \equiv 
 \frac{1}{4 \pi}  \int_{-\infty}^t dt'  {\bf E} \cdot \frac{\partial {\bf D}}{\partial t'}   
=\frac{1}{8 \pi} \int_{-\infty}^{\infty} d\omega \int_{-\infty}^{\infty} d\omega'  
\left[{\om'\eps^*(\om')-\om\eps(\om)\over\om'-\om}\right] {\bf E}({\bf r},\omega) \cdot {\bf E}^*({\bf r},\omega')
 e^{i (\omega'-\omega) t} ,
\label{poynt1}
\ee
\be
\mathcal{W}_M(\br,t) \equiv 
 \frac{1}{4 \pi}  \int_{-\infty}^t dt'  {\bf H} \cdot \frac{\partial {\bf B}}{\partial t'}=
\frac{1}{8 \pi} \int_{-\infty}^{\infty} d\omega \int_{-\infty}^{\infty} d\omega'  
\left[{\om'\mu^*(\om')-\om\mu(\om)\over\om'-\om}\right] {\bf H}({\bf r},\omega) \cdot {\bf H}^*({\bf r},\omega')
 e^{i (\omega'-\omega) t},
\label{poynt2}
\ee
\end{widetext}
obtained using the properties (\ref{RefRels}). (To simplify notation henceforth we do not indicate 
any dependence of $\eps$ and $\mu$ on $\br$.) The constants of integration vanish under the assumption that ${\bf E}({\bf r}, t) \rightarrow 0$ and ${\bf H}({\bf r}, t) \rightarrow 0$ as $t \rightarrow - \infty$.

So far we have been using the concept of electromagnetic energy density a bit loosely, and at this point we would like to make our statements more precise. Being a direct consequence of Maxwell's equations, the balance relation (\ref{Poynting2}) is valid under arbitrary thermodynamical conditions, so it can be used to describe general out-of-equilibrium systems. However, in those situations one should be careful on defining energies, as it is easy to see that in such systems the r.h.s. of (\ref{Poynting2}) contains not only the ``standard'' electromagnetic energy $W(\br,t)$ but also the dissipated heat $Q(\br,t)$, and, as it is discussed in length by Barash and Ginzburg \cite{Barash1976, GinzburgPlasma}, in general it is impossible to separate the two in a unambiguous way. So, in order to avoid confusion, we always work with the sum 
\be
\mathcal{W}(\br,t) = W(\br,t) + Q(\br,t),
\ee
and we refer to $\mathcal{W}(\br,t)$ as the electromagnetic energy for the lack of a better term, but always bearing the above considerations in mind. In the case where thermal equilibrium is established, the evolved heat $Q(\br,t)$ vanishes and $\mathcal{W}(\br,t)$ coincides with $W(\br,t)$. 
In addition, even for situations in thermal equilibrium, we shall make the distinction of the EM energy when there is absorption present ($\epsilon_I, \mu_I \neq 0$), that we
will call $W({\bf r},t)$, and when absorption is absent ($\epsilon_I=\mu_I =0$), that we will denote by $u({\bf r},t)$.

In the limiting case of a monochromatic field we have
\be
\bbE(\br,\om)={1\over 2}\bbE_{\om_0}(\br)[\delta(\om-\om_0)+\delta(\om+\om_0)] ,
\ee
and, if absorption at frequency $\om_0$ is negligible, it follows from (\ref{poynt1}) that at thermal equilibrium
\be
\overline{u}_E(\br,\om_0)=\frac{1}{16 \pi}\left[\frac{d (\omega \epsilon_R)}{d \omega} \right]_{\om_0} |{\bf E}_{\om_0}({\bf r})|^2,
\ee
where we have averaged over the period $2\pi/\om_0$ and have used
\be
{\om'\eps^*(\om')-\om\eps(\om)\over\om'-\om}={\om'\eps_R(\om')-\om\eps_R(\om)\over\om'-\om}\rightarrow
{d\over d\om}(\om\eps_R) 
\ee
for $\eps(\om)=\eps_R(\om)$ and $\om'\rightarrow\om$. Together with the corresponding result for $u_M(\br)$, this
gives Eq. (\ref{NoAbsorption1}). 


\subsection{Quasi-monochromatic fields}

In the case of quasi-monochromatic fields it is convenient to write the fields ${\bf E}({\bf r},t)$ and 
${\bf H}({\bf r},t)$ as
\begin{eqnarray}
\label{quasi-monocromatic}
{\bf E}({\bf r},t)&=& \frac{1}{2} \left[ {\bf E}_0({\bf r},t) e^{-i\omega_0 t} + {\bf E}_0^*({\bf r},t) e^{i\omega_0 t} \right]  \nonumber \\
&=&\int_{0}^{\infty}d\omega[{\bf E}({\bf r},\omega) e^{-i\omega t}+{\bf E}^*({\bf r},\omega) e^{i\omega t}],\\
{\bf H}({\bf r},t) &=& \frac{1}{2} \left[ {\bf H}_0({\bf r},t) e^{-i\omega_0 t} + {\bf H}_0^*({\bf r},t) e^{i\omega_0 t}\right]  \nonumber \\ 
&=& \int_{0}^{\infty}d\omega[{\bf H}({\bf r},\omega) e^{-i\omega t}+{\bf H}^*({\bf r},\omega) e^{i\omega t}], \nonumber \\
\end{eqnarray}
where the envelope functions ${\bf E}_0({\bf r},t)$ and ${\bf H}_0({\bf r},t)$ 
vary slowly in time compared to $e^{-i\omega_0  t}$. The Fourier components ${\bf E}({\bf r},\omega)$ and 
${\bf H}({\bf r},\omega)$ in this case are sharply peaked at the frequency $\omega_0$, and we consider the time-averages 
($\overline{\mathcal{W}}_E$ and $\overline{\mathcal{W}}_H$) of $\mathcal{W}_E$ and $\mathcal{W}_H$ over times long compared to $2 \pi / \omega_0$ but
short compared to times over which ${\bf E}_0$ and ${\bf H}_0$ vary significantly. We also assume
that $\epsilon(\omega)$ and $\mu(\omega)$ vary slowly enough near $\om=\om_0$ so that
we can retain only the first-order terms in their Taylor series about $\omega_0$ \cite{Barash1976}:
\begin{eqnarray}
\label{TaylorEpsMu}
\omega \epsilon(\omega) \hspace{-8pt}&&\cong \omega_0 \epsilon(\omega_0) + \frac{d (\omega \epsilon)}{d\omega}\bigg|_{\omega_0} \hspace{-7pt}(\omega - \omega_0)  \nonumber \\
 \hspace{-8pt}&&= \omega \epsilon(\omega_0) + \frac{d \epsilon}{d\omega}\bigg|_{\omega_0} \hspace{-7pt}(\omega - \omega_0), \nonumber \\
\omega \mu(\omega) \hspace{-8pt}&&\cong \omega_0 \mu(\omega_0) + \frac{d (\omega \mu)}{d\omega}\bigg|_{\omega_0} \hspace{-7pt}(\omega - \omega_0)  \nonumber \\
 \hspace{-8pt}&&= \omega \mu(\omega_0) + \frac{d \mu}{d\omega}\bigg|_{\omega_0} \hspace{-7pt}(\omega - \omega_0) .
\end{eqnarray}  
With these approximations we find straightforwardly that 
\begin{eqnarray}
\label{AvEnApproxE}
\overline{\mathcal{W}}_E(t) &\cong&  \frac{1}{16 \pi} 
\frac{d (\omega \epsilon_R)}{d \omega}\bigg|_{\omega_0}  |{\bf E}_0({\bf r},t)|^2   \nonumber \\ 
&&\mbox{} + {\omega_0t\over 8\pi} \epsilon_I (\omega_0)|{\bf E}_0({\bf r},t)|^2 ,
\end{eqnarray}
and
\begin{eqnarray}
\label{AvEnApproxH}
\overline{\mathcal{W}}_M(t) &\cong& \frac{1}{16 \pi}  
\frac{d (\omega \mu_R)}{d \omega}\bigg|_{\omega_0}  |{\bf H}_0({\bf r},t)|^2 \nonumber \\ 
&&\mbox{} \hspace{25pt} +  {\omega_0t\over 8\pi} \mu_I (\omega_0)|{\bf H}_0({\bf r},t)|^2,
\end{eqnarray}
where we have used 
\begin{eqnarray}
\label{AveField}
\overline{{\bf E}(\br,t)^2}&=&\int_{0}^{\infty} \!\!\! d\omega \int_{0}^{\infty} \!\!\! d\omega'  \, {\bf E}^*({\bf r},\omega) \cdot {\bf E}({\bf r},\omega') e^{i (\omega - \omega')t}  \nonumber \\
&=&{1\over 2}|{\bf E}_0(\br,t)|^2 ,\nonumber \\
\overline{{\bf H}(\br,t)^2}&=&\int_{0}^{\infty} \!\!\! d\omega \int_{0}^{\infty} \!\!\! d\omega'  \, {\bf H}^*({\bf r},\omega) \cdot {\bf H}({\bf r},\omega') e^{i (\omega - \omega')t} \nonumber \\
&=& {1\over 2}|{\bf H}_0(\br,t)|^2.
\end{eqnarray}
The time $t$ here has been assumed to be short compared to the time over which the slowly varying
field envelopes ${\bf E}_0(\br,t)$ and ${\bf H}_0(\br,t)$ change significantly, as otherwise even small deviations from monochromaticity can invalidate the approximations (\ref{AvEnApproxE}) and (\ref{AvEnApproxH}). 

Equation (\ref{AvEnApproxE}) gives the first two terms corresponding to the expression (8) of Barash and 
Ginzburg \cite{Barash1976}. To obtain the remaining terms in that expression we must include terms proportional to $d\epsilon_I/d\omega$, that may be written as
\bea
&&\frac{1}{16\pi}\frac{i \,d(\omega \epsilon_I)}{d \omega} \int_{-\infty}^{\infty} \!\!\! d\omega \! \int_{-\infty}^{\infty} \!\!\! d\omega'  \frac{(\om - \om_0)+(\om' - \om_0)}{\om' -\om} \nonumber \\
&& \hspace{20pt} \times \; {\bf E}({\bf r},\omega) \cdot {\bf E}^*({\bf r},\omega')e^{i (\omega'-\omega) t} ,
\eea
and, after straightforward manipulations, as
\bea
\frac{1}{16\pi}\frac{i \,d(\omega \epsilon_I)}{d \omega} \hspace{-12pt}&&\int_{-\infty}^t \!\!dt' \left[ \frac{d {\bf E}_0(\br,t')}{d t'}  {\bf E}_0^*(\br,t') \right. \nonumber \\
&&\left. - \frac{d {\bf E}_0^*(\br,t')}{d t'}  {\bf E}_0(\br,t') \right] .
\eea 
which is nothing but the time integral of the third term in Eq. (8) of \cite{Barash1976}. Since the discussion of quasi-monochromatic fields is not the main goal of this paper we shall stop it here, referring the reader to \cite{Barash1976,GinzburgPlasma} for further details. There one can find extensive discussions on the interplay of EM energy and heat generated, on how dissipation makes it (in general) impossible to write the ``standard'' EM energy $W(\br,t)$ in terms of dielectric functions alone, how it is possible to go beyond the quasi-monochromatic approximation when dissipation is absent, etc.


\subsection{Uncorrelated frequencies}

Another situation where it is possible to simplify the general expression for the energy density is when
we have stochastic fields such that their autocorrelation functions in the frequency domain satisfy
\begin{eqnarray}
\label{CorrFourier1}
\langle\langle {\bf E}({\bf r},\omega) \cdot {\bf E}^*({\bf r},\omega') \rangle\rangle = \frac{1}{2} |{\bf E}({\bf r},\omega)|^2 \delta(\omega' - \omega),\nonumber \\
\langle\langle  {\bf H}({\bf r},\omega) \cdot  {\bf H}^*({\bf r},\omega') \rangle\rangle = \frac{1}{2} |{\bf H}({\bf r},\omega)|^2 \delta(\omega' - \omega) , \nonumber \\
\langle\langle {\bf E}({\bf r},\omega) \cdot {\bf E}({\bf r},\omega') \rangle\rangle = \langle\langle {\bf H}({\bf r},\omega) \cdot {\bf H}({\bf r},\omega') \rangle\rangle = 0,
\end{eqnarray}
where $\langle\langle ...\rangle\rangle$ denotes the average over the appropriate ensemble. From these correlation
functions it follows that correlations in the time domain are stationary:
\begin{eqnarray}
\label{Correlations}
\langle\langle {\bf E}({\bf r},t) \cdot {\bf E}({\bf r},t') \rangle\rangle&=&F_E(\br,t-t') ,\nonumber \\
\langle\langle {\bf H}({\bf r},t) \cdot {\bf H}({\bf r},t') \rangle\rangle &=& F_H(\br,t-t') .
\end{eqnarray}
It is clear that equations (\ref{Correlations}) are satisfied when there is no net dissipation or gain, which means that either $\epsilon_I = \mu_I = 0$ or that there are Langevin-type forces in the system that compensate for dissipated energy. Restricting ourselves to the first case in this simple example, we can use (\ref{CorrFourier1}) to calculate the ensemble average of (\ref{poynt1}) and (\ref{poynt2}) and obtain at thermal equilibrium
\begin{eqnarray}
\label{AvEnEnsemble}
\langle\langle {u}_E + {u}_M \rangle\rangle &=& 
 {1\over 16\pi}\int_0^{\infty}d\om \left[ {d\over d\om}(\om\eps_R) \langle\langle|{\bf E}({\bf r},\omega)|^2\rangle\rangle \right. \nonumber \\
&&\mbox{} + \left. {d\over d\om}(\om\mu_R) \langle\langle|{\bf H}({\bf r},\omega)|^2\rangle\rangle \right] .
\end{eqnarray}
Of course this expression is strictly valid only over frequency ranges at which absorption is negligible; in such
ranges the integrand of Eq. (\ref{AvEnEnsemble}) gives {\it exactly} the spectral energy density.


\subsection{Classical oscillator model for the energy density}

In order to better focus on the physics involved in these considerations of energy density we briefly review a 
classical model \cite{loudon} in which the medium consists of $N$ harmonic oscillators per unit volume, each having a natural oscillation frequency $\Om$ and satisfying the equation of motion
\be
\ddot{\bbx}+\Om^2{\bbx}={e\over m}\bbE .
\label{loudon1}
\ee
(For notational simplicity we do not indicate here the $\br$ dependence of $\bbE$.)
The polarization density and dielectric constant in this model are respectively
\be
{\bf P}=Ne\bbx={Ne^2/m\over\Om^2-\om^2} ,
\label{loudon2}
\ee
and
\be
\eps_R(\om)=1+{4\pi Ne^2/m\over\Om^2-\om^2}=1-{\om_p^2\over\om^2-\Om^2} ,
\label{loudon3}
\ee
where $\om_p=(4\pi Ne^2/m)^{1/2}$ is the plasma frequency. We write Poynting's theorem in its integral form:
\bea
\oint{\bf S}\cdot\hat{n}da&=&-{1\over 4\pi}\int\left[\bbE\cdot{\pa{\bbD}\over\pa t}+\bbH\cdot{\pa{\bbH}\over\pa t}\right]dV
\nonumber \\
&=&-{1\over 4\pi}\int\left[{1\over 2}{\pa\over\pa t}(\bbE^2+\bbH^2)+4\pi\bbE\cdot{\pa{\bf P}\over\pa t}\right]dV .
\nonumber
\label{loudon4}
\eea
The integral of the normal component of ${\bf S}$ on the left-hand
side is, as usual, over a surface enclosing a volume $V$. From (\ref{loudon1}),
\be
\bbE\cdot{\pa{\bf P}\over\pa t}={m\over e}(\ddot\bbx+\Om^2\bbx)\cdot Ne\dot{\bbx}=N{\pa\over\pa t}
\left({1\over 2}m\dot{\bbx}^2+{1\over 2}m\Om^2\bbx^2\right)  ,
\label{loudon5}
\ee
and therefore 
\be
\oint{\bf S}\cdot\hat{n}da=-\int\dot{u}dV ,
\label{loudon6}
\ee
\be
u\equiv {1\over 8\pi}(\bbE^2+\bbH^2)+N\left({1\over 2}m\dot{\bbx}^2
+{1\over 2}m\Om^2\bbx^2\right)  .
\label{loudon7}
\ee
$u$ is the density of total energy, that in the field plus that in the medium. Using 
\be
\bbx={e/m\over\Om^2-\om^2}\bbE_{\om}\cos\om t \ \ \ \ \ \ {\rm and} \ \ \ \ \ \dot{\bbx}=-{\om e/m\over\Om^2-
\om^2}\bbE_{\om}\sin\om t \ ,
\label{loudon8}
\ee
and (\ref{loudon3}) for a monochromatic field $\bbE_{\om}\cos\om t$, we find after cycle-averaging that
\bea
\overline{u}&=&{1\over 16\pi}\bbE_{\om}^2+{1\over 16\pi}\bbH_{\om}^2+{Ne^2\over 4m}{\Om^2+\om^2\over(\Om^2-\om^2)^2}\bbE_{\om}^2 \nonumber \\
&=&{1\over 16\pi}\left[1+{\om_p^2\over\Om^2-\om^2}+{2\om^2\om_p^2\over(\Om^2-\om^2)^2}\right]\bbE_{\om}^2
+{1\over 16\pi}\bbH_{\om}^2 \nonumber \\
&=&{1\over 16\pi}\left[\eps_R(\om)+\om{d\eps_R\over d\om}\right]\bbE_{\om}^2+{1\over 16\pi}\bbH_{\om}^2 ,
\label{loudon88}
\eea
confirming that equation (\ref{NoAbsorption1}) defines the {\sl total} energy density of a dielectric medium
($\mu_R=1$). From the relation $\bbH_{\om}^2=\eps_R(\om)\bbE_{\om}^2$, 
\be
\overline{u}={1\over 8\pi}\left[\eps_R+{1\over 2}\om{d\eps_R\over d\om}\right]\bbE_{\om}^2,
\ee
the term  $(1/16\pi)\om[d\eps_R/d\om]\bbE_{\om}^2$ is seen from (\ref{loudon7}) and (\ref{loudon8}) to be the (cycle-averaged) {\it kinetic} energy per unit volume of the material oscillators in this model.

Absorption is included in this model by adding $\Gamma\dot{\bbx}$ ($\Gamma>0$) to the left-hand side of 
(\ref{loudon1}) \cite{loudon}.
Then it is easily shown that the rate of change of energy density $\mathcal{W}$ in the volume $V$, defined such 
that 
\be
\int_V{\pa \mathcal{W}\over\pa t}dV=-\oint{\bf S}\cdot\hat{n}da,
\ee
is
\bea
{\pa \mathcal{W}\over\pa t}&=&N{\pa\over\pa t}\left[{1\over 2}m\dot{\bbx}^2+{1\over 2}m\Om^2\bbx^2\right]\nonumber \\
&&\mbox{}+{\pa\over\pa t}\left[{1\over 8\pi}(\bbE^2+\bbH^2)\right]
+2\Gamma N({1\over 2}m\dot{\bbx}^2).
\label{classenergy}
\eea
A similar expression is derived in the QED theory in Section \ref{sec:micro}. An important difference, however,
is that in the QED theory there is an additional term arising from Langevin forces, which are required for the
preservation of commutation relations. In addition, we should stress that these Langevin forces also ensure thermal equilibrium even when dissipation is present, and in fact all our discussion about the quantum case is restricted to systems in thermal equilibrium. Of course we can also include a Langevin force in a classical model
in order to balance dissipative effects and obtain an average energy consistent with thermal equilibrium.


\section{QED energy density in a uniform, dispersive, non-absorbing medium in thermal equilibrium}\label{sec:zpe}

One major distinction between the classical and QED theories, of course, is that in QED there is
a nonvanishing zero-point energy associated among other things with Casimir effects. For purposes of comparison
with results obtained in the following section when absorption is included, we reproduce here the thermal equilibrium QED
energy density in a uniform, dispersive medium in which absorption is negligible. To do this we simply 
regard Eqs. (\ref{poynt1}) and (\ref{poynt2}) as expectation values when expressed in symmetrized form in terms of the quantized fields $\bE$ and $\bH$. (We  use  circumflexes to designate operators.) For the zero-temperature (vacuum) state $|0\ra$, for example, we use familiar expectation values, e.g., 
\bea
\la 0|\bE^{(+)}(\br,\om)\cdot\bE^{(-)}(\br,\om')|0\ra&=&|\bbE(\br,\om)|^2\delta(\omega-\omega'),\nonumber \\
\la 0|\bE^{(-)}(\br,\om)\cdot\bE^{(+)}(\br,\om')|0\ra&=&0,
\eea
where $\bE^{(+)}(\br,\om)$ and $\bE^{(-)}(\br,\om)$ are respectively the photon annihilation and creation parts
of $\bE(\br,\om)$, and obtain straightforwardly
\begin{widetext}
\begin{equation}
\frac{1}{8 \pi}  \int_{-\infty}^t \!\! dt'  \langle 0 |\hat{{\bf E}} \cdot \frac{\partial \hat{{\bf D}}}{\partial t'} + \frac{\partial \hat{{\bf D}}}{\partial t'} \cdot \hat{{\bf E}}| 0 \rangle = 
\frac{1}{4 \pi}  \int_{0}^{\infty} d\omega 
\frac{d (\omega \epsilon_R)}{d \omega} |{\bf E}({\bf r},\omega)|^2 , 
\end{equation}
and
\begin{equation}
\frac{1}{8 \pi}  \int_{-\infty}^t \!\! dt' \langle 0 |\hat{{\bf H}} \cdot \frac{\partial \hat{{\bf B}}}{\partial t'} + \frac{\partial \hat{{\bf B}}}{\partial t'} \cdot \hat{{\bf H}}| 0 \rangle = 
\frac{1}{4 \pi}     \int_{0}^{\infty} d\omega 
\frac{d (\omega \mu_R)}{d \omega} |{\bf H}({\bf r},\omega)|^2 ,
\end{equation}
\end{widetext}
by manipulations similar to those used in the preceding section.
For the case of a uniform medium with negligible absorption, these expressions imply the (infinite) zero-point energy density
\begin{eqnarray}
u_E+u_M \hspace{-10pt}&&={1\over 8\pi}\sum_{\lambda}\int_0^{\infty}d\om 
\left[ 
{d\over d\om}(\om\eps_R)E_{\lambda\om}^2 \right. \nonumber \\
&&\left. + {d\over d\om}(\om\mu_R)H_{\lambda\om}^2
\right] ,
\label{class}
\end{eqnarray}
where $\lambda=1,2$ denotes polarization components. The zero-point squared amplitudes for the
quantized field in a non-absorbing medium are \cite{pwmdiel}
\begin{eqnarray}
&& E^2_{\lambda\om}= \frac{\hbar}{\pi c^3}   \mu_R(\omega) n_R(\omega)  \omega^3 ,\nonumber \\
&& H^2_{\lambda\om}= \frac{\hbar}{\pi c^3}  \frac{1}{\mu_R(\omega)} n_R^3(\omega) \omega^3, 
\end{eqnarray}
where $n_R(\om)= (\eps_R(\om) \mu_R(\om))^{1/2}$ is the refractive index. Therefore we have the familiar
result
\begin{eqnarray}
\label{zpe}
u_E+u_M \hspace{-10pt}&&= 
{\hbar\over 2\pi^2c^3}\int_0^{\infty}d\om\om^3n^2_R(\om){d\over d\om}[\om n_R(\om)]  \nonumber \\ 
&& = {\hbar\over 2\pi^2c^2}\int_0^{\infty} d\om\om^3n^2_R(\om){1\over v_g(\om)}  \nonumber \\
&&= {\hbar\over 2\pi^2}\int_0^{\infty}dkk^2\om \nonumber \\
&&= {2\over 8\pi^3}\int d^3k{1\over 2}\hbar\om ,
\end{eqnarray}
where we have used the relation $k=n_R(\om)\om/c$ and the definition  
$v_g(\om)=c[d(n_R \om)/d\om]^{-1}$ of the group velocity at frequency $\om$.
Whereas we have obtained this result for the zero-point energy
without taking absorption into account, we will show in the following section that it is valid in general for a
uniform absorbing medium.

The generalization to finite temperatures is similarly
straightforward and yields, of course,
\bea u_E+u_M&=&{2\over 8\pi^3}\int d^3k\left[{1\over 2}\hbar\om+{\hbar\om\over e^{\hbar\om/k_BT}-1}\right] \nonumber \\
&\equiv& \int_0^{\infty}d\om\rho(\om),
\label{classT}
\eea
where the spectral energy density \cite{ginzbook,pwmdiel}
\be
\rho(\om)={n_R^2(\om)\hbar\om^3\over \pi^2v_g(\om)c^2}\left({1\over 2}+{1\over e^{\hbar\om/k_BT}-1}\right).
\ee


\section{Model for a dispersive, absorbing dielectric medium in thermal equilibrium}\label{sec:micro}

Following the work of many others \cite{manyothers}, we model a dielectric medium as a collection of harmonic oscillators. Aside from the need to introduce oscillator strengths in order to obtain correct numerical results, the oscillator model is an excellent approximation if the atoms of a dielectric medium remain with high probability in their ground states. Each oscillator atom has a mass $m$ and a natural frequency $\om_0$ and is coupled to a 
reservoir of other harmonic oscillators responsible for the damping of its oscillations and homogeneous line broadening of its (electric-dipole) transition. The Hamiltonian
for this model, including the electromagnetic field and its coupling to the atoms, is
\bea
\hat{H}&=&{1\over 8\pi}\int d{\bf r}(\bE^2+\bH^2)+\sum_j \left( {1\over 2m}[\bp_j-{e\over c}\bA(\br_j)]^2 \right. \nonumber \\
&&\left. \mbox{}+{1\over 2}m\om_0^2\bx_j^2 \right)
+\int_0^{\infty}d\om\hbar\om\sum_j\left[\bb^{\dag}_j(\om)\cdot\bb_j(\om)+{1\over 2}\right] \nonumber \\
&&\mbox{}-i\int_0^{\infty}d\om \Lambda(\om)\sum_j\bx_j\cdot[\bb_j(\om)-\bb_j^{\dag}(\om)] .
\label{eq1} 
\eea
The first two terms correspond in standard notation to the energy of the electromagnetic field, the atom oscillators,
and their coupling via the vector potential $\bA(\br_j)$, where $\br_j$ denotes the position of the $j$th
atom. The third and fourth terms represent respectively the energy of the reservoir oscillators and their 
interaction with the atoms. The reservoir oscillators satisfy the bosonic commutation relations
\be
[\hat{b}_{i\mu}(\om),\hat{b}_{j\nu}^{\dag}(\om')]=\delta_{ij}\delta_{\mu\nu}\delta(\om-\om') , \ \ \ 
[\hat{b}_{i\mu}(\om),\hat{b}_{j\nu}(\om')]=0 \ ,
\label{eq2}
\ee
where we use Greek letters to denote Cartesian components of vectors. The atom-reservoir coupling constant is chosen to be
\be
\Lambda(\om)=\left({m\hbar\gamma\om\over\pi}\right)^{1/2}
\label{eq3}
\ee
in order that each atom's oscillations be damped at the rate $\gamma$, as shown below.  

From (\ref{eq2}) and $[\hat{x}_{i\mu},\hat{p}_{j\nu}]=i\hbar\delta_{ij}\delta_{\mu\nu}$ we obtain in the dipole approximation the Heisenberg 
equations of motion
\be
{\ddot{\bx}}_j+\om_0^2\bx_j={e\over m}\bE(\br_j)+{i\over m}\int_0^{\infty}d\om \Lambda(\om)[\bb_j(\om,t)-\bb_j^{\dag}(\om,t)] ,
\label{eq4}
\ee
\be
{\dot{\bb}}_j(\om,t)=-i\om \bb_j(\om,t)+{1\over\hbar}\Lambda(\om)\bx_j \ .
\label{eq5}
\ee
Using the formal solution of (\ref{eq5}) in (\ref{eq4}), it follows that
\bea
{\ddot{\bx}}_j+\om_0^2\bx_j&=&{e\over m}\bE(\br_j)+{1\over m}\bF_{Lj}(t)+{i\over m\hbar} \nonumber \\
&\times& \hspace{-7pt} \int_0^{\infty}d\om \Lambda^2(\om)
\int_0^tdt'\hat{\bf x}_j(t')[2i\sin\om(t'-t)], \nonumber \\
\label{eq6}
\eea
where the ``Langevin force" operator $\bF_{Lj}(t)$ acting on the $j$th atom is 
\be
\bF_{Lj}(t)=i\int_0^{\infty}d\om \Lambda(\om)[\bb_j(\om,0)e^{-i\om t}-\bb_j^{\dag}(\om,0)e^{i\om t}] .
\label{lang}
\ee
The third term on the right-hand side of (\ref{eq6}) is
\bea
&-&{2\over m\hbar}\int_0^{\infty}d\om \Lambda^2(\om)\int_0^tdt'\bx_j(t')\sin\om(t'-t) \nonumber \\
&=&-{2\gamma\over\pi}\int_0^tdt'\bx_j(t')
\int_0^{\infty}d\om\om\sin\om(t'-t) \nonumber \\
&=&{2\gamma}\int_0^tdt'\bx_j(t'){\partial\over\partial t'}\delta(t'-t)
=-\gamma{\dot{\bx}}_j(t) \ .
\label{damp}
\eea
We have omitted a divergent frequency shift which, when the atom-reservoir coupling is modified by a form factor 
to produce a finite expression, can be assumed to be included in the definition of the atom's transition 
frequency $\om_0$. Then (\ref{eq6}) simplifies to a ``quantum Langevin equation" \cite{fordkac}:
\be
{\ddot{\bx}}_j+\gamma{\dot{\bx}}_j+\om_0^2\bx_j={e\over m}\bE(\br_j)+{1\over m}\bF_{Lj}(t)  .
\label{eq7}
\ee

In the absence of coupling to the electromagnetic field we have, for times $t\gg\gamma^{-1}$,
\bea
\bp_j(t)&=&m{\dot{\bx}}_j(t)=\int_0^{\infty}d\om\om \Lambda^2(\om)\left[{\bb_j(\om)e^{-i\om t}\over\om_0^2-\om^2-i\gamma\om} \right.
\nonumber \\
&&\left.+{\bb_j^{\dag}(\om)e^{i\om t}\over\om_0^2-\om^2+i\gamma\om}\right] .
\label{eq8}
\eea
(We now write $\bb_j(\om)$ in place of $\bb_j(\om,0)$.) Similarly, using (\ref{eq2}), we obtain \cite{ford1}
\bea
&&[\hat{x}_{i\mu}(t),\hat{p}_{j\nu}(t')]=\delta_{ij}\delta_{\mu\nu}{2i\hbar\gamma\over\pi}\int_0^{\infty}
{d\om\om^2\cos\om(t'-t)\over(\om_0^2-\om^2)^2
+\gamma^2\om^2} \nonumber \\
&=&i\hbar\delta_{ij}\delta_{\mu\nu}\left[\cos\om_1(t'-t)-{\gamma\over 2\om_1}\sin\om_1|t'-t|\right]e^{-\gamma|t'-t|/2} , \nonumber \\
\label{eq9}
\eea
where $\om_1\equiv[\om_0^2-\gamma^2/4]^{1/2}$; thus the canonical commutation relation at equal times $[\hat{x}_{i\mu}(t),\hat{p}_{j\nu}(t)]=i\hbar\delta_{ij}\delta_{\mu\nu}$ is preserved in the coupling of the atom to the 
reservoir \cite{gardiner}. 

The energy expectation value of a single oscillator without coupling to the electromagnetic field is 
found similarly to be
\bea
\la{1\over 2}m{\dot{\bx}}_j^2+{1\over 2}m\om_0^2\bx_j^2\ra&=&{\hbar\gamma\over 2\pi}\int_0^{\infty}d\om
{\om(\om_0^2+\om^2)\over(\om_0^2-\om^2)^2+\gamma^2\om^2} \nonumber \\
&&\mbox{}\times \sum_{\mu=1}^3[2\la \hat{b}^{\dag}_{j\mu}(\om)\hat{b}_{j\mu}(\om)\ra +1]. \nonumber \\
\label{eq10}
\eea
Since we are working in the Heisenberg picture, the expectation value is over the initial state of the
coupled system of oscillators. If we assume that the reservoir is in an initial state of thermal equilibrium at temperature $T$, while the oscillator coupled to it is in its ground state, then
\bea
\la \hat{b}^{\dag}_{i\mu}(\om)\hat{b}_{j\nu}(\om')\ra&=&\la \hat{b}_{i\mu}(\om)\hat{b}^{\dag}_{j\nu}(\om')\ra-
\delta_{ij}\delta_{\mu\nu}\delta(\om-\om')
\nonumber \\
&=&{1\over e^{\hbar\om/k_BT}-1}\delta_{ij}\delta_{\mu\nu}\delta(\om-\om')\nonumber \\
&\equiv& {\cal N}(\omega) \delta_{ij}\delta_{\mu\nu}\delta(\om-\om')
\label{btherm}
\eea
and
\bea
\la{1\over 2}m{\dot{\bx}}_j^2+{1\over 2}m\om_0^2\bx_j^2\ra&=&{3\hbar\gamma\over\pi}\int_0^{\infty}d\om
{\om(\om_0^2+\om^2){\cal N}(\om)\over(\om_0^2-\om^2)^2+\gamma^2\om^2} \nonumber \\
&+&{3\hbar\gamma\over 2\pi}\int_0^{\infty}d\om{\om(\om_0^2+\om^2)\over(\om_0^2-\om^2)^2+\gamma^2\om^2}. \nonumber \\
\eea
The first term on the right is just the energy of the oscillator in thermal equilibrium, and has a rather complicated
closed form \cite{ford2}; it becomes just  $3\hbar\om_0/[e^{\hbar\om_0/k_BT}-1]$ in the weak-coupling 
limit ($\gamma\rightarrow 0$). The second term is the zero-point energy of the oscillator, that for $\om_1 > 0$ may be written as
\be
{3\hbar\over\pi}\om_1\cos^{-1}\left({\gamma\over 2\om_0}\right)+{3\hbar\gamma\over 2\pi}\ln\left({\om_c\over\om_0}\right),
\ee
where $\om_c$ is a high-frequency cutoff \cite{ford3}. It reduces to $(3\hbar\om_0)/2$ in the weak-coupling limit. 


\subsection{Noise Polarization}

The Heisenberg equations of motion for the electric and magnetic fields that follow from the Hamiltonian (\ref{eq1}) have, of course, the same form as their classical (Maxwell) counterparts
\begin{eqnarray}
\label{MaxEqs}
&&\nabla \times \hat{\bf E} = - \frac{1}{c}\frac{\partial\hat{\bf B}}{\partial t} , \nonumber \\
&&\nabla \times \hat{\bf H} = \frac{4 \pi}{c} \hat{{\bf J}} + \frac{1}{c}\frac{\partial\hat{\bf E}}{\partial t}  ,
\end{eqnarray}
that must be supplemented with
\bea
\label{MaxEqs2}
 \nabla \cdot \hat{\bf B} =  0 \;\;\; , \;\;\; \nabla \cdot \hat{\bf D} =  0 , 
\eea
where
\begin{eqnarray}
&& \hat{\bf D} =  \hat{\bf E} + 4\pi\hat{\bf P}, \nonumber \\
&&\hat{\bf J}({\bf r},t) = \frac{\partial\hat{\bf P}({\bf r},t)}{\partial t}, \nonumber \\
&&\hat{\bf P}({\bf r},t)= e\sum_j\bx_j(t)\delta^3(\br-\br_j),
\end{eqnarray}
and, because our model does not induce any magnetic activity, we have $\hat{\bf B} =  \hat{\bf H}$. It is advantageous to work in the frequency domain, so we write
\bea
\bE(\br,t)&=&\int_0^{\infty}d\om[\bE(\br,\om)e^{-i\om t}+\bE^{\dag}(\br,\om)e^{i\om t}]  , \nonumber \\
\bH(\br,t)&=&\int_0^{\infty}d\om[\bH(\br,\om)e^{-i\om t}+\bH^{\dag}(\br,\om)e^{i\om t}]  , \nonumber \\
\bP(\br,t)&=&\int_0^{\infty}d\om[\bP(\br,\om)e^{-i\om t}+\bP^{\dag}(\br,\om)e^{i\om t}]  ,
\label{eq11}
\eea
where the Fourier transform of the polarization density may be written as
\bea
\bP(\br,\om)&=&e\sum_j\bx_j(\om)\delta^3(\br-\br_j), 
\label{eq12a}\\
\bx_j(t)&=&\int_0^{\infty}d\om[\bx_j(\om)e^{-i\om t}+\bx_j^{\dag}(\om)e^{i\om t}] \ .
\label{eq13}
\eea
It follows from (\ref{eq7}) that
\bea
\bP(\br,\om)&=&{e^2/m\over\om_0^2-\om^2-i\gamma\om}\sum_j\bE(\br_j,\om)\delta^3(\br-\br_j) \nonumber \\
&&\mbox{}+{ie/m\over \om_0^2-\om^2-i\gamma\om}\Lambda(\om)\sum_j\bb_j(\om)\delta^3(\br-\br_j) \nonumber \\
&\rightarrow&{Ne^2/m\over\om_0^2-\om^2-i\gamma\om}\bE(\br,\om) \nonumber \\
&+&{iNe/m\over \om_0^2-\om^2-i\gamma\om}\Lambda_c(\om)\bb(\br,\om) , 
\label{eq15}
\eea
in the approximation in which we assume the atoms are continuously distributed with a density $N$ and $\Lambda_c(\omega) =  \sqrt{\rho_m \hbar\gamma\omega / \pi}$, with $\rho_m = m/N$.

From Maxwell's equations (\ref{MaxEqs}, \ref{MaxEqs2}) and the fact that $\nabla \cdot \bb(\br,\om) = 0$ \cite{divb}, we obtain
\bea
&&\nabla^2\bE(\br,\om)+{\om^2\over c^2}\bE(\br,\om)=-4\pi{\om^2\over c^2}\bP(\br,\om) \nonumber \\
&=&{-4\pi Ne^2/m\over\om_0^2-\om^2-i\gamma\om}{\om^2\over c^2}\bE(\br,\om)- \nonumber \\
&&\mbox{}{4\pi ieN/m\over \om_0^2-\om^2-i\gamma\om}{\om^2\over c^2}\Lambda(\om)\bb(\br,\om) , \nonumber \\
\label{eq16}
\eea
or
\be
\nabla^2\bE(\br,\om)+{\om^2\over c^2}\eps(\om)\bE(\br,\om)=-{\om^2\over c^2}\bK(\br,\om) ,
\label{eq17}
\ee
where the complex permittivity is
\bea
\eps(\om)&=&1-{4\pi Ne^2/m\over\om^2-\om_0^2 + i\gamma\om}\equiv 1- {\om_p^2\over\om^2-\om_0^2 + i\gamma\om} \nonumber \\
&=&\eps_R(\om)+i\eps_I(\om).
\label{eq18}
\eea
We have also defined the ``noise polarization" at frequency $\om$:
\be
\bK(\br,\om)={4\pi iNe/m\over\om_0^2-\om^2-i\gamma\om}\Lambda(\om)\bb(\br,\om) .
\label{noisepol}
\ee
This contribution to the polarization arises from the Langevin force $\bF_{Lj}(t)$ in the quantum Langevin 
equation (\ref{eq7}). Its principal properties for our purposes are the thermal equilibrium expectation values
\bea 
&&\la \hK_{\mu}(\br,\om)\ra = \la \hK^{\dag}_{\mu}(\br,\om)\ra = 0 , \nonumber \\
&&\hspace{-15pt}\la \hK_{\mu}(\br,\om)\hK_{\nu}(\br',\om')\ra = \la \hK^{\dag}_{\mu}(\br,\om)\hK^{\dag}_{\nu}(\br',\om')\ra = 0 ,
\eea 
and
\bea
\la\hK^{\dag}_{\mu}(\br,\om)\hK_{\nu}(\br',\om')\ra&=&4\hbar\eps_I(\om)\delta_{\mu\nu}\delta(\om-\om')\delta^3(\br-\br')
\nonumber \\
&&\mbox{}\times{1\over e^{\hbar\om/k_BT}-1},
\label{prop1}
\eea
\bea
\la\hK_{\mu}(\br,\om)\hK^{\dag}_{\nu}(\br',\om')\ra&=&4\hbar\eps_I(\om)\delta_{\mu\nu}\delta(\om-\om')\delta^3(\br-\br')
\nonumber \\
&&\mbox{}\times \left[ {1\over e^{\hbar\om/k_BT}-1}+1 \right] ,
\label{prop2}
\eea
all of which follow from (\ref{btherm}) and $\la \hat{b}_{i\mu}(\om)\hat{b}_{j\nu}(\om')\ra=0$. Eqs. (\ref{prop1}, \ref{prop2}) constitute nothing else than the fluctuation-dissipation theorem, that we {\it derived} from the fundamental assumptions of a canonical bath and a linear coupling to the matter. 

We can proceed formally now as in Lifshitz's paper  \cite{lifshitz} and define operators $\hg_{\lambda}(\bk,\om)$ by writing
\be
\hat{\bf K}(\br,\om)=\int d^3k\sum_{\lambda=1,2}\hg_{\lambda}(\bk,\om)\bek e^{i\bk\cdot\br}.
\label{eqk}
\ee
{ The solenoidal character of $\bb(\br,\om)$ implies directly in $\nabla\cdot\hat{\bf K}(\br,\om)=0$} and therefore we can choose the vectors $\bek$ such 
that $\bk\cdot\bek=0$, $\bek\cdot{\bf e}_{\bk\lambda'}=0$, $\lambda=1,2$; we also take the $\bek$ to be real. Then
\bea
\hg_{\lambda}(\bk,\om)&=&\left({1\over 2\pi}\right)^3\int d^3r\,\hat{\bf K}(\br,\om)
\cdot\bek e^{-i\bk\cdot\br} \nonumber \\
&\equiv&\left({1\over 2\pi}\right)^3\int d^3r\hK_{\lambda}(\br,\om) e^{-i\bk\cdot\br} ,
\eea
and Eqs. (\ref{noisepol}) and (\ref{eq2}) imply the commutation relation 
\bea
[\hg_{\lambda}(\bk,\om),\hg^{\dag}_{\lambda'}(\bk',\om') ] &=&{\hbar\over 2\pi^3}\eps_I(\om)\delta_{\lambda\lambda'}
\delta(\om-\om')\nonumber \\
&&\mbox{}\times\delta^3(\bk-\bk') ,
\eea
where again we make the uniform continuum approximation for the spatial distribution of the material oscillators.
Finally it will be convenient to introduce the operators 
\be
\hC_{\lambda}(\bk,\om)\equiv [\hbar\eps_I(\om)/2\pi^3]^{-1/2}\hg_{\lambda}(\bk,\om) ,
\label{eqkk}
\ee
satisfying 
\be
[\hC_{\lambda}(\bk,\om),\hC_{\lambda'}^{\dag}(\bk',\om')]=\delta_{\lambda\lambda'}\delta(\om-\om')\delta^3(\bk-\bk').
\label{eqC}
\ee

\subsection{Remarks on the Lifshitz Theory}
It seems worthwhile as an aside to compare the formulation presented thus far with the Lifshitz theory. For this 
purpose we write 
\bea
&&{1\over 2}\la\hK^{\dag}_{\mu}(\br,\om)\hK_{\nu}(\br',\om')+\hK_{\mu}(\br,\om)\hK^{\dag}_{\nu}(\br',\om')\ra \nonumber \\
&=&4\hbar\eps_I(\om)\delta_{\mu\nu}\delta(\om-\om')\delta^3(\br-\br')
\left[ {1\over e^{\hbar\om/k_BT}-1}+{1\over 2} \right] . \nonumber \\
\label{ourfluct}
\eea
The right-hand side is equivalent to that in equation (1.2) of Lifshitz's paper \cite{lifshitz}, which in our notation has the form
\bea
\langle \langle K_{\mu}^*(\br,\om)K_{\nu}(\br',\om') \rangle \rangle&=&4\hbar\eps_I(\om)\delta_{\mu\nu}\delta(\om-\om')\delta^3(\br-\br') \nonumber \\
&&\mbox{}\times \left[ {1\over e^{\hbar\om/k_BT}-1}+{1\over 2} \right] ,
\label{rytov}
\eea
the $\langle \langle \ldots \rangle \rangle$ denoting a {\it classical} ensemble average. This expression in the Lifshitz theory is a statement
of a fluctuation-dissipation relation that Lifshitz attributes to Rytov \cite{lifshitz}. The difference between
(\ref{ourfluct}) and (\ref{rytov}) reflects the fact that in the Lifshitz theory the thermal equilibrium electric, magnetic, and noise polarization fields are treated in effect as classical fluctuating 
fields; the fluctuation-dissipation relation (\ref{rytov}) is used to relate the average of the square of the noise polarization to the imaginary part of the permittivity. The constant $\hbar$ appears in (\ref{rytov}) as a result of
fixing the right-hand side such that the average over the classical ensemble for the squared fields matches the
corresponding quantum expectation values. In our (quantum) formulation based on the quantum Langevin
equation for the material oscillators, the only nonvanishing contribution to the 
expectation value of the square of the noise polarization at $T=0$, for instance, is 
\be
\la\hK_{\mu}(\br,\om)\hK^{\dag}_{\nu}(\br',\om')\ra=4\hbar\eps_I(\om)\delta_{\mu\nu}\delta(\om-\om')\delta^3(\br-\br'),
\ee
twice the corresponding result in Lifshitz's paper. But Lifshitz's averages for the squared fields
are the same as our corresponding quantum expectation values because in his formulation both 
$\langle \langle K^*(\br,\om)K(\br',\om') \rangle \rangle$ and $\langle \langle K(\br,\om)K^*(\br',\om') \rangle \rangle$ contribute (equally) to these
averages. Thus the same Casimir force will be obtained in
either approach because they both involve the same average zero-point energy per mode, although
of course the averages in the two approaches are fundamentally different. In the Lifshitz theory, in which
forces between bodies are calculated using the stress tensor, there are
no quantized fields, and averages of components of the stress tensor are over classical ensembles of stochastic fields, their statistical properties being determined by imposing the Rytov fluctuation-dissipation 
relation (\ref{rytov}).

For the calculation of the Casimir force between perfectly conducting plates, for example, a stochastic 
electrodynamical (SED) theory yields the correct force when $\hbar$ is introduced by requiring that there is a zero-point field energy $(1/2)\hbar\om$ per mode of frequency $\om$ \cite{boyer}. In SED, as in the Lifshitz theory, both 
${\bf E}^*(\br,\om)\cdot{\bf E}(\br,\om)$ and ${\bf E}(\br,\om)\cdot{\bf E}^*(\br,\om)$ contribute to the average of the squared electric field at $T=0$, whereas in our quantized-field approach only $\bE(\br,\om)\cdot\bE^{\dag}(\br,\om)$ contributes. The Lifshitz approach to the calculation of Casimir forces may be regarded as an application SED 
in which dissipation as well as finite thermal equilibrium temperatures are treated.

\subsection{Electric and Magnetic Fields}

An expression for the quantized electric field in an absorptive dielectric now follows directly from
Eqs. (\ref{eq11}), (\ref{eq17}), (\ref{eqk}), and (\ref{eqkk}):
\bea
\bE(\br,t)&=&\int d^3k\sum_{\lambda}\int_0^{\infty}d\om\sqrt{\hbar\eps_I(\om)/2\pi^3}{\om^2/c^2\over k^2-
\eps(\om)\om^2/c^2}\nonumber \\
&&\mbox{}\times\hC_{\lambda}(\bk,\om)
\bek e^{-i(\om t-\bk\cdot\br)}+{\rm h.c.} .
\label{eq33}
\eea
From $\nabla\times\bE=
-(1/c)\pa\bB/\pa t$ we also obtain an expression for the magnetic field (reminding that $\mu=1$ under our assumption in this 
section of a nonmagnetic medium):
\bea
\bH(\br,t)&=&ic\int d^3k\sum_{\lambda}\int_0^{\infty}d\om\sqrt{\hbar\eps_I(\om)/2\pi^3}\nonumber \\
&&\mbox{}\times{\om^2/c^2\over k^2-\eps(\om)\om^2/c^2}\hC_{\lambda}(\bk,\om)\nonumber \\
&&\mbox{}\times\om^{-1}\left( \bk \times \bek\right) e^{-i(\om t-\bk\cdot\br)} +{\rm h.c.} .
\label{eq34}
\eea
These expressions have the same form as the corresponding ones obtained by Huttner and Barnett \cite{huttnerbarnett} by Fano diagonalization of the entire system of coupled harmonic oscillators.

One quantity of interest is the zero-temperature expectation value of $\bE^2(\br,t)$, for which the considerations above yield
\bea
\la\bE^2(\br,t)\ra&=&{\hbar\over 2\pi^3c^4}\int_0^{\infty}d\om\int d^3k\sum_{\lambda}{\om^4\eps_I(\om)\over
|k^2-\eps(\om)\om^2/c^2|^2}\nonumber \\
&=&{\hbar\over 2\pi^3c^4}\sum_{\lambda}\int_0^{\infty}d\om\om^4\eps_I(\om)\nonumber \\
&&\mbox{}\times\int_0^{\infty}{4\pi k^2dk\over{[k^2-\om^2\eps_R(\om)/c^2]^2+\om^4\eps_I^2(\om)/c^4}}\nonumber \\
&=&{\hbar\over\pi c^3}\sum_{\lambda}\int_0^{\infty}d\om\om^3n_R(\om),
\nonumber \\
\label{elecspec}
\eea
where we have used the relations $\eps_R(\om)=n_R^2(\om)-n_I^2(\om)$ and $\eps_I(\om)=2n_R(\om)n_I(\om)$
for the real and imaginary parts ($n_R$ and $n_I$) of the refractive index. We note that this is
the same form one would obtain by quantizing the field in a dispersive and {\sl non-absorbing} medium,
assuming a purely real permittivity $\eps_R(\om)$ \cite{pwmdiel}.

For $\bH^2(\br,t)$ we obtain the zero-point expectation value
\bea
\la\bH^2(\br,t)\ra&=&{\hbar\over 2\pi^3c^2}\sum_{\lambda}\int_0^{\infty}d\om\om^2\eps_I(\om)\nonumber \\
&&\mbox{}\times\int_0^{\infty}{4\pi k^4dk\over|k^2-\eps(\om)\om^2/c^2|^2}.\nonumber \\
&=& {\hbar\over 2\pi^3c^2} {\rm Im} \sum_{\lambda}\int_0^{\infty}d\om\om^2 \eps(\om) \nonumber \\
&&\mbox{} \times \int \frac{d^3 k}{k^2 - \om^2 \eps(\om)/c^2}
\label{regul}
\eea
The integral over $k$ diverges. However, it is obtained in the approximation that the atoms of the dielectric form
a continuum, an approximation that is invalid when $ka\gg 1$, where $a$ is a typical interatomic spacing.
To apply our macroscopic approach based on the characterization of the medium by a permittivity 
$\eps(\om)$---a long-photon-wavelength approximation implicit in the Lifshitz theory---we must ``regularize"
the integral (\ref{regul}) to extract a finite result. In this case it is convenient to introduce a Lorentzian cutoff to the $k$-integral \cite{Barnett} and then, by taking advantage of the integrand even parity, a simple application of the residue theorem gives (in the continuum limit)
\be
\label{DivergentInt}
\lim_{a \rightarrow 0} \int {d^3k \over k^2-\eps\om^2/c^2}\frac{1}{1+k^2a^2}= \frac{2\pi^2}{a} + 2\pi^2  i {\om\over c}\eps^{1/2} ,
\ee
where, for clarity, we omitted the $\om$-dependence of $\eps(\om)$. Finally, a direct substitution of (\ref{DivergentInt}) into (\ref{regul}) gives
\bea
\frac{1}{8\pi}\la\bH^2(\br,t)\ra={\hbar\over 8 \pi^2 c^3}\sum_{\lambda}\int_0^{\infty}\!\!\!d\om\om^2 \left[\frac{\epsilon_I c}{a} + \om\, {\rm Re} \, \epsilon^{3/2} \right] \! , \nonumber \\
\label{magspec}
\eea
which, of course, diverges as $a \rightarrow 0$ even before the $\omega$-integration. We shall see in the next subsection that the divergent part is actually canceled out when all the contributions for the energy are taken into account.


\subsection{Energy Density}

To obtain an expression for the total energy density in the dielectric medium we start from
Poynting's theorem in the conventional notation (symmetrized Poynting operator
$\hat{\bf S}=(c/8\pi)[\bE\times\bH-\bH\times\bE]$) and take expectation values
over the initial state of the system consisting of the field, the dielectric atoms, and the reservoir:
\bea
\label{Poynting3}
\oint\la\hat{\bf S}\ra\cdot{\bf n}da&=&-{1\over 8\pi}\int\la\bE\cdot{\pa\bD\over\pa t}+{\pa\bD\over\pa t}
\cdot\bE\ra dV
\nonumber \\
\mbox{}&&-{1\over 8\pi}\int\la\bH\cdot{\pa\bH\over\pa t}+{\pa\bH\over\pa t}\cdot\bH\ra dV.
\eea
According to the usual interpretation, the l.h.s.  of (\ref{Poynting3}) is the energy flux through a given surface $S$ and, given that we are assuming thermal equilibrium within our system, it should vanish. Thermal equilibrium also allows us to identify the rate of change with time of the expectation value of the total energy $W$ per unit volume:
\be
{\pa W\over\pa t}={1\over 8\pi}\la\bE\cdot{\pa\bD\over\pa t}+{\pa\bD\over\pa t}\cdot\bE\ra 
+{1\over 8\pi}{\pa\over\pa t}\la\bH^2\ra ,
\ee
which, of course, also vanishes. For the system under consideration $\bD=\bE+4\pi\bP_{\eps}+\bK$, where $\bP_{\eps}$ is the part of the
polarization giving rise to the dielectric permittivity $\eps(\om)$ and $\bK$ is the noise polarization defined by
(\ref{noisepol}). Thus $\bD=\bD_{\eps}+\bK$ and
\be
{\pa W\over\pa t}={\pa W_1\over\pa t}+{\pa W_2\over\pa t},
\ee
where we define
\be
{\pa W_1\over\pa t}={1\over 8\pi}\la\bE\cdot{\pa\bD_{\eps}\over\pa t}+{\pa\bD_{\eps}\over\pa t}\cdot\bE\ra
+{1\over 8\pi}{\pa\over\pa t}\la\bH^2\ra
\ee
and
\be
\label{W2}
{\pa W_2\over\pa t}={1\over 8\pi}\la \bE\cdot{\pa\bK\over\pa t}+{\pa\bK\over\pa t}\cdot\bE\ra.
\ee

Before proceeding with the calculation of $W$ we note the following identity that follows from our model of the dielectric:
\bea
{\pa W\over \pa t}&=&\la{\pa\over\pa t}\sum_j\left[{1\over 2}m{\dot{\bx}}_j^2+{1\over 2}m\om_0^2{\bx_j}^2\right]\delta^3(\br-\br_j)
\nonumber \\
&&\mbox{}+{1\over 4\pi}{\pa\over\pa t}\left[\bE^2+\bH^2\right]\nonumber \\
&&\mbox{}+\sum_j[2\gamma({1\over 2}m{\dot{\bx}}_j^2)-{\dot{\bx}}_j\cdot{\bf F}_{Lj}]\delta^3(\br-\br_j)\ra.\nonumber \\
\label{phys}
\eea
The first term is the rate of change of the energy density (kinetic plus potential) of the oscillators 
constituting the dielectric, and the second term is the rate of change of the energy density of the
electromagnetic field. In the absence of any dissipation ($\gamma=0$ and therefore ${\bf F}_{Lj}=0$),
the third term on the right vanishes, and $W=u$ is just the {\sl total} (matter-plus-field) energy density. 
The third term accounts for the effect of the reservoir on the dielectric oscillators: $2\gamma\sum_j({1\over 2}m{\dot{\bx}}_j^2)\delta^3(\br-\br_j)$
is the rate of change of kinetic energy density due to the dissipative effect of the reservoir, while
$\sum_j{\dot{\bx}}_j\cdot{\bf F}_{Lj}\delta^3(\br-\br_j)$ is the rate of work per unit volume done by the
Langevin forces on the dielectric oscillators. In the absence of the electromagnetic interaction these
effects cancel, and the third term in (\ref{phys}) again vanishes. There is a close formal similarity between (\ref{phys}) and the corrresponding expression
(\ref{classenergy}) that follows from the classical oscillator model. The essential physical difference 
between (\ref{phys}) and (\ref{classenergy}) lies simply in the effect of the Langevin force term in the quantum-electrodynamical expression of energy conservation.

To obtain the total energy density we focus first on the case of zero temperature, as the result for 
finite temperature requires only a simple extension of the zero-temperature calculation,
as discussed below. Using (\ref{eq11}) plus
\bea
{\pa\bD_{\eps}\over\pa t}\hspace{-7pt}&&=-i\int_0^{\infty}d\om\om[\eps(\om)\bE(\br,\om)e^{-i\om t} \nonumber \\
&& \hspace{20pt} - \eps^*(\om)\bE^{\dag}(\br,\om)e^{+i\om t}],
\eea
and integrating over $t$, we obtain
\bea
W_1(\br,t)&=&{1\over 8\pi}\sum_{\lambda} \int_0^{\infty}d\om'\int_0^{\infty}d\om
\frac{\om'\eps^*(\om')-\om\eps(\om)}{\om'-\om}\nonumber \\
&&\mbox{}\times \la\bE_{\lambda}(\br,\om)\cdot\bE^{\dag}_{\lambda}(\br,\om')\ra e^{-i(\om-\om')t}
\nonumber \\
&&\mbox{}+{1\over 8\pi}\la\bH^2(\br,t)\ra,
\label{w11}
\eea
where we have used the fact that the vacuum (zero-temperature) expectation value 
$\la\bE^{\dag}_{\lambda}(\br,\om)\cdot\bE_{\lambda'}(\br,\om')\ra=0$ while
 $\la\bE_{\lambda}(\br,\om)\cdot\bE^{\dag}_{\lambda'}(\br,\om')\ra$ vanishes unless $\lambda=\lambda'$
and $\om=\om'$. To deal with what appears to be a singularity at $\om=\om'$ we rewrite (\ref{w11})
as a sum of two identical terms and interchange $\om$ and $\om'$ in the second one, to get
\begin{widetext}
\bea
\label{w1-2}
W_1(\br,t)&=&{1\over 8\pi}\sum_{\lambda} \int_0^{\infty}\!d\om'\int_0^{\infty}\!d\om
\frac{\om'\eps_R(\om')-\om\eps_R(\om)}{\om'-\om} \la\bE_{\lambda}(\br,\om)\cdot\bE^{\dag}_{\lambda}(\br,\om')\ra e^{-i(\om-\om')t} \nonumber \\
&&- {i\over 8\pi}\sum_{\lambda}\! \int_0^{\infty}\!\!\!\!d\om'\int_0^{\infty}\!\!\!\!d\om \left(\om'\eps_I(\om')+\om\eps_I(\om)\right) \frac{\la\bE_{\lambda}(\br,\om)\cdot\bE^{\dag}_{\lambda}(\br,\om')\ra e^{-i(\om-\om')t} - \la\bE_{\lambda}(\br,\om')\cdot\bE^{\dag}_{\lambda}(\br,\om)\ra e^{i(\om-\om')t}}{2(\om'-\om)} \nonumber \\
&&+{1\over 8\pi}\la\bH^2(\br,t)\ra .
\eea
Next we use (\ref{eq33}) and (\ref{eqC}) to write the vacuum expectation value
\bea
\la\bE_{\lambda}(\br,\om)\cdot\bE^{\dag}_{\lambda}(\br,\om')\ra = \la\bE_{\lambda}(\br,\om')\cdot\bE^{\dag}_{\lambda}(\br,\om)\ra = {\hbar\over 2\pi^3}\eps_I(\om){\om^4\over c^4} \int d^3k{1\over |k^2-\eps(\om)\om^2/c^2|^2} \delta(\om-\om'),
\eea
that allows us to readily evaluate the first term in (\ref{w1-2}) by noticing that
\bea
\lim_{\om'\rightarrow\om}{\om'\eps_R(\om')-\om\eps_R(\om)\over\om'-\om}={d\over d\om}[\om\eps_R(\om)].
\eea 
The second term is calculated by realizing that the zeroth order contributions in $(\om-\om')$ in the numerator cancel each other, while the first order terms produce a contribution linear in the elapsed time $t$
\bea
\lim_{\om'\rightarrow\om}\frac{e^{-i(\om-\om')t} - e^{i(\om-\om')t}}{2(\om'-\om)}=it .
\eea
Therefore
\bea
W_1(\br,t)&=&\!{1\over 8\pi}{\hbar\over 2\pi^3c^4}\sum_{\lambda} \int_0^{\infty}\!\!\!d\om \!\left({d\over d\om}[\om\eps_R] + 2t\omega\eps_I \!\right) \!\om^4
\eps_I \int d^3k{1\over |k^2-\eps\om^2/c^2|^2}+{1\over 8\pi}\la\bH^2(\br,t)\ra\nonumber \\
&=&{\hbar\over 8\pi^2c^3}\sum_{\lambda}\int_0^{\infty}\!\!\!d\om\om^3n_R {d\over d\om}[\om\eps_R]  +{1\over 8\pi}\la\bH^2(\br,t)\ra \, + \, t \cdot {\hbar\over 4\pi^2c^3}\sum_{\lambda}\int_0^{\infty}\!\!\!d\om\om^4 n_R \eps_I  
\label{w1}
\eea
\end{widetext}
where we have again used the relations $\eps_R=n_R^2-n_I^2$ and $\eps_I=2n_Rn_I$ and we are leaving the $\om$-dependence implicit in both $\eps$ and $n$. Let us note that the rate of change in time of $W_1(t)$ is a positive constant, given by the last term in (\ref{w1})\cite{footNr}. This implies that this term is responsible for creating heat in any given volume $V$ (meaning an inward flux of energy). Since we know that in thermal equilibrium the total flux should vanish, this energy increase must be balanced out by an energy decrease coming from $W_2(t)$.

To evaluate $W_2$ as given in Eq. (\ref{W2}) we first define $\bK(\bk,\om)$ by writing
\bea
\label{KTot}
\bK(\br,t)&=&\int_0^{\infty}d\om\int d^3k\sum_{\lambda}[\bK_{\lambda}(\bk,\om)e^{-i\om t}e^{i\bk\cdot\br}\nonumber \\
&&\mbox{}+\bK^{\dag}_{\lambda}(\bk,\om)e^{i\om t}e^{-i\bk\cdot\br}],
\eea
and use (\ref{eqk}), (\ref{eqkk}), and (\ref{eq33}) to relate $\bK_{\lambda}(\bk,\om)$ and $\bE_{\lambda}(\bk,\om)$:
\be
\label{Klambda}
\bK_{\lambda}(\bk,\om)={c^2\over\om^2}[k^2-\eps(\om)\om^2/c^2]\bE_{\lambda}(\bk,\om).
\ee
Then, after inserting (\ref{KTot}) and (\ref{Klambda}) in (\ref{W2}) and a few algebraic steps, we get 
\begin{widetext}
\bea
W_2(\br,t)&=&- \frac{\hbar}{16\pi^4 c^2} \sum_{\lambda} \int_0^{\infty}\!\!\! d\om' \int_0^{\infty} \!\!\! d\om \frac{\om^2\om'}{\om-\om'} \sqrt{\eps_I(\om)\eps_I(\om')} \delta(\om-\om') \int d^3k \left[ \frac{e^{-i(\om-\om')t}}{k^2-\eps(\om)\om^2/c^2} + \frac{e^{i(\om-\om')t}}{k^2-\eps^*(\om)\om^2/c^2} \right],\nonumber \\
\eea
and, proceeding as in the evaluation of $W_1$, we obtain
\bea
W_2(\br,t)&=& - {\hbar\over 8\pi^4 c^2}\sum_{\lambda}{\rm Re}\int_0^{\infty} \!\!\! d\om'\int_0^{\infty}\!\!\! d\om {\om^2\om'\over\om-\om'}\sqrt{\eps_I(\om)\eps_I(\om')} \delta(\om-\om')\int d^3k{1\over k^2-\eps(\om')\om'^2/c^2} \nonumber \\
&&- \; t \cdot {\hbar\over 4\pi^2c^3}\sum_{\lambda}\int_0^{\infty}\!\!\!d\om\om^4 n_R \, \eps_I  ,
\label{w2-2}
\eea
\end{widetext}
where we have used the integral $\int d^3 k |k^2 - \eps \om^2/c^2|^{-2} = 2 \pi^2 c n_R / \eps_I \om$ (as in Eq. (\ref{elecspec})) to obtain the second term. Now we see clearly that the time dependent term in $W_2(t)$ precisely cancels the one in $W_1(t)$, ensuring thermal equilibrium. Let us note also that the first term in (\ref{w2-2}) contains the same $k$-integral as the one present in (\ref{regul}), and therefore we may invoke Eq. (\ref{DivergentInt}) to evaluate it. The first term of (\ref{w2-2}) is then equal to
\be
-\frac{\hbar}{8\pi^4 c^2} \sum_{\lambda} \int_0^{\infty}\!\!\! d\om' \! \int_0^{\infty} \!\!\! d\om \frac{\om^2\om'}{\om-\om'}\sqrt{\eps_I \eps_I'} \, \frac{2 \pi^2}{a} \, \delta(\om-\om') ,
\ee  
where we again left implicit the $\om$- and $\om'$-dependences in $\eps_I$. The apparent singularity in the $\om' \rightarrow \om$ limit may be dealt with by using the procedure described just before Eq. (\ref{w1-2}), and then after some trivial steps we get
 \be
- {\hbar\over 8 \pi^2 c^2}\sum_{\lambda}\int_0^{\infty}d\om\om^2 \frac{\epsilon_I}{a} ,
 \ee
what cancels exactly the first term in (\ref{magspec}). We still have to work on the contribution of the second term of (\ref{DivergentInt}) to the first term of (\ref{w2-2}), that leads to 
\bea
&&{\hbar\over 8\pi^2c^3}\sum_{\lambda}{\rm Im}\int_0^{\infty}d\om\int_0^{\infty}d\om'\nonumber \\
&&\hspace{-25pt}\times \! \lim_{\om'\rightarrow\om}
{\om\om'\sqrt{\eps_I(\om)\eps_I(\om')}\over\om-\om'}[\om^2\eps^{1/2}(\om)
-\om'^2\eps^{1/2}(\om')] \nonumber \\
&&\hspace{-25pt}={\hbar\over 8\pi^2c^3}\sum_{\lambda}{\rm Im}\int_0^{\infty}\!\!d\om\om^2\eps_I(\om){d\over d\om}[\om^2\eps^{1/2}(\om)].
\label{w2}
\eea
The total energy density is obtained by adding (\ref{w1}) and (\ref{w2-2}) and using (\ref{w2}), (\ref{magspec}):
\bea
W&=&{\hbar\over 8\pi^2c^3}\sum_{\lambda}\int_0^{\infty}d\om\om^3 \bigg\{ {\rm Re}\left[n_R{d\over d\om}(\om\eps)+\eps^{3/2}\right]\nonumber \\
&&\mbox{}+{1\over\om}\eps_I{\rm Im}{d\over d\om}(\om^2\eps^{1/2})\bigg\}.
\label{energyabsorption}
\eea
The above expression of the energy density of a uniform, dispersive and absorbing medium, is the most important result of this paper.
Using $\eps(\om)=n^2(\om)$ and the following relations
\bea
n_R{d\over d\om}(\om\eps_R) &&\hspace{-8pt}= (n_R^2-n_I^2)n_R  \nonumber \\
&&\hspace{-8pt} + \, \om n_R \left( 2n_R \frac{dn_R}{d\om} - 2n_I \frac{dn_I}{d\om} \right) , \nonumber \\
{\rm Re} \, \eps^{3/2} &&\hspace{-8pt}= (n_R^2-n_I^2)n_R - 2n_Rn_I^2 , \nonumber \\
\frac{\eps_I}{\om} {\rm Im} \frac{d}{d\om} (\om^2 \sqrt{\eps}) &&\hspace{-8pt}= 4n_Rn_I^2 + 2n_Rn_I \om \frac{dn_I}{d\omega} ,
\eea
and summing over polarizations we obtain our final expression for the vacuum expectation value of the total energy density:
\bea
\label{FinalExpression}
W=&&\hspace{-8pt}{\hbar\over 2\pi^2c^3}\int_0^{\infty}d\om\om^3n^2_R(\om)\left( n_R + \om {d n_R \over d\om} \right) \nonumber \\
=&&\hspace{-8pt}{\hbar\over 2\pi^2c^3}\int_0^{\infty}d\om\om^3n^2_R(\om){d\over d\om}[\om n_R(\om)],
\eea
which is just (\ref{zpe}): the QED zero-point energy density depends in exactly the same way on the refractive index,
regardless of whether absorption is accounted for, and in fact it depends only on the real part of the refractive index.

The same conclusion holds for finite temperatures. In this case both $\la\bE^{\dag}_{\lambda}(\br,\om)\cdot\bE_{\lambda'}(\br,\om')\ra$ and
 $\la\bE_{\lambda}(\br,\om)\cdot\bE^{\dag}_{\lambda'}(\br,\om')\ra$ make nonvanishing contributions to $W_1$ and
 $W_2$. Using 
\bea
&&\la\hC_{\lambda}^{\dag}(\bk,\om)\hC_{\lambda'}(\bk',\om')+
\hC_{\lambda}(\bk,\om)\hC^{\dag}_{\lambda'}(\bk',\om')\ra \nonumber \\
&&\mbox{}=\delta_{\lambda,\lambda'}\delta(\om-\om')\delta^3(\bk-\bk')\coth{{\hbar\om\over 2k_BT}},
\eea
we arrive straightforwardly at exactly the formula (\ref{classT}) for the total energy density in thermal equilibrium. 


\section{Concluding Remarks}\label{sec:summ}
We have shown that the approximate classical expression (\ref{NoAbsorption1}) for the energy density in a band of frequencies at which absorption in a dielectric can be ignored is in fact {\sl exactly} correct as a spectral average value in (i) classical theory in the case where the fields at different frequencies within the band undergo uncorrelated fluctuations, and (ii) QED at zero temperature or more generally at thermal equilibirum.

Using the model of a dielectric medium as a collection of harmonic oscillators, and including the
coupling of each oscillator to a reservoir of oscillators that give rise to dissipation and a Langevin force
on each oscillator, we have shown how a noise polarization results from these reservoirs 
and compared it with that employed in Lifshitz's well-known theory of Casimir effects. From this comparison we
concluded that the Lifshitz theory is actually a classical stochastic electrodynamical theory. We arrived at quantized
electric and magnetic fields having the same form as in the Huttner-Barnett approach in which the complete Hamiltonian
is diagonalized, and showed that the expectation value of the total energy of the system of dielectric oscillators, reservoirs, and the electromagnetic field has the same form in thermal equilibrium, including the limiting case of zero temperature, independent of whether we take dissipation into account in quantizing the field. 

Our treatment allowed us to derive the celebrated fluctuation-dissipation theorem, and also to show explicitly that it ensures that in thermal equilibrium the total energy of the system of oscillators, reservoir and electromagnetic field is constant in time. When absorption is present, there is a coupling between the system of oscillators and the reservoir and an energy exchange between them. In our example of dielectric medium modeled by a collection of harmonic oscillators, we have explicitly shown that a positive energy rate $\dot{W}_1 > 0$ arising from the interaction of the EM field with the system is exactly canceled by a corresponding negative energy rate coming from the interaction of the system with the reservoir, $\dot{W}_2 = -\dot{W}_1 < 0$.  We should stress that this energy rate balance is absolutely general, and applies not only to absorbing dielectric media, as treated here, but also to arbitrary dissipative materials, e.g. metals modeled by a dissipative Drude permittivity. This is merely a manifestation of the fluctuation-dissipation theorem, and as such holds for Lifshitz theory, which assumes thermal equilibrium.

These conclusions cannot be applied directly to the question raised in the Introduction: why do derivations of Casimir forces that start from calculations of changes in zero-point energy of {\it presumed} dissipationless media appear in the end to produce the same results as in the Lifshitz theory, where a correlation function of a fluctuating 
(``noise") polarization is related to the imaginary (dissipative) part of the permittivity? To address this question we must take into account a hallmark of Casimir effects, namely the role of boundaries. We have restricted ourselves here to the model of perfectly uniform media with no boundaries; the role of boundaries will be addressed in a
forthcoming paper \cite{paperII}.

Finally we mention two examples, not involving boundaries, where these conclusions are consistent with known
results. The first example is spontaneous emission of an atom embedded in a dielectric medium. The Einstein
$A$ coefficient for the rate of spontaneous emission at the electric dipole transition frequency $\om_0$ is proportional to the zero-temperature expectation value of $\bE^2(\br,\om_0)$, which from (\ref{elecspec}) is seen to be proportional to $n_R(\om_0)$. Therefore the $A$ coefficient for an atom in a dielectric medium with complex refractive index $n_R(\om)+
in_I(\om)$ is just $n_R(\om_0)$ times the free-space $A$ coefficient \cite{barloud1}. This assumes the
continuum approximation for the dielectric; near-field interactions of the embedded excited atom with host atoms,
including local field corrections, result in a rate of energy loss by the excited atom that depend on $n_I(\om_0)$
\cite{otherspon}. 

The second example, less straightforard but more closely related to Casimir forces, is the van der Waals interaction between two atoms embedded in a dielectric medium. It has been shown that the interactions between electrically or magnetically polarizable atoms can be obtained from the quantized electric and magnetic fields in a nonabsorbing medium \cite{vdwdiel}. Absorption affects the final expression for the interaction only after the permittivity (or permeability) is properly regarded as a complex function of frequency, analytic in the upper half of the complex frequency plane. 
As in the Lifshitz theory, and as in the present work, the calculations leading to this result are based on the continuum model of the dielectric medium.

\section*{Acknowledgement}
We thank S.M. Barnett, L.S. Brown, S.Y. Buhmann, I.E. Dzyaloshinskii, J.H. Eberly and R.F. O'Connell for helpful comments relating to this research. This work was funded by DARPA/MTO's Casimir Effect Enhancement program under DOE/NNSA Contract DE-AC52-06NA25396.



\begin{thebibliography}{99}
\bibitem{divb} This is just the statement that there is no "bath flow" anywhere ($\oint \bb \cdot da = 0$), what is very natural assumption in a homogeneous material (its permittivity is independent of position).
\bibitem{Barnett} We are greatly indebted to S.M. Barnett's comments on this issue, in particular for pointing out the advantages of a Lorentzian cutoff.
\bibitem{footNr} It could be argued that this term is not necessarily positive since it contains the factor $n_R$ (that may be negative for certain types of materials), but for non-magnetic materials one can show that indeed $n_R > 0 \;\forall\, \om$ by the following reasoning: since we are assuming $\eps_I > 0$, we have $0 < {\rm Arg} \, \eps < \pi$ and therefore $0 < {\rm Arg} \, n = {\rm Arg} \sqrt{\eps} < \pi/2$, which proves that $n_R > 0$.
\end{thebibliography}
\end{document}